# Scintillation properties of ceramics based on zinc oxide

Vladimir A. Demidenko[a], Elena I. Gorokhova[a], Ivan V. Khodyuk[b,*], Ol'ga A. Khristich[a], Sergey B. Mikhrin[b], Piotr A. Rodnyi[b]

[a]*Federal State Unitary Organization "Research and Technological Institute of Optical Materials All-Russia Scientific Center", S.I. Vavilov State Optical Institute, Babushkina 36-1, 192171 Saint-Petersburg, Russia*
[b]*St. Petersburg State Polytechnic University, Polytekhnicheskaya 29, 195251 Saint-Petersburg, Russia*

**Abstract**

Ceramics ZnO:Zn of 20 mm diameter and 1.6 mm thickness with an optical transparency up to 0.33 in the visible region have been prepared by hot pressing technique. Scintillating and luminescent characteristics such as emission spectra, decay time, yield, and TSL glow curve have been measured under X-ray excitation. Two emission bands peaking at 500 and 380 nm were detected, the light output was about 80% of that for standard BGO scintillator, main decay constant was $10.4 \pm 0.1$ ns. The obtained data allow us to consider the ZnO:Zn ceramics as a perspective scintillator. Finally, the investigation shows that other ZnO-based fast scintillators can be fabricated in the form of optical ceramics.
© 2007 Elsevier Ltd. All rights reserved.

## 1. Introduction

The well known blue-green phosphor ZnO:Zn, that is zinc oxide containing excess zinc, has been under investigation for a few decades (Shionoya and Yen, 1999), because of its use in vacuum fluorescent and field-emission displays. Luminescent characteristics of the ZnO:Zn are studied in detail though physical mechanisms responsible for its emission are barely understood yet (Shionoya and Yen, 1999; Van Dijken et al., 2000). Zinc oxide is non-hygroscopic, stable over a wide range of temperatures, mechanically robust, and sufficiently dense, 5.61 g/cm$^3$. Currently some compounds based on ZnO are considered as promising fast scintillators (Moses, 2002; Simpson et al., 2003; Kubota et al., 2004). Efficient ultrafast ($\tau < 1$ ns) scintillations have been observed in ZnO:Ga and ZnO:In phosphors (Moses, 2002; Simpson et al., 2003). The ZnO:Zn scintillators containing $^6$Li have been developed for high-counting-rate neutron imaging (Kubota et al., 2004; Katagiri et al., 2004). The scintillators based on ZnO show a high radiation hardness and appropriate stopping power. Such scintillating materials could hold a central position among fast scintillators, however, for many purposes the large-volume scintillators are required. Zinc oxide is available mostly in powder or thin film form and only recently a small single crystal ZnO has been produced (Simpson et al., 2003).

In this work, we have studied some scintillating and thermo-stimulate luminescent (TSL) properties of the ZnO:Zn ceramics.

## 2. Experimental

A hot pressing technique was used to transform the ZnO:Zn powder into a transparent ceramic scintillator. A problem when sintering ZnO ceramics is the hexagonal symmetry of its crystal structure. Anisotropic grains in the ceramics cause considerable light scattering in the grain boundaries resulting in light absorption in the bulk. Nevertheless, the ZnO:Zn ceramics with reasonable optical transparency (Fig. 1) have been obtained. To create a high quality optical ZnO:Zn ceramics, we have used our preceding experience in the ceramics sintering from non-cubic materials (see e.g., Gorokhova et al., 2005). Inset of Fig.1 illustrates a good optical transparency of the ZnO:Zn ceramics.

---

* Corresponding author. Tel.: +7 812 555 3739; fax: +7 812 552 7574.
*E-mail address:* khodyuk@list.ru (I.V. Khodyuk).

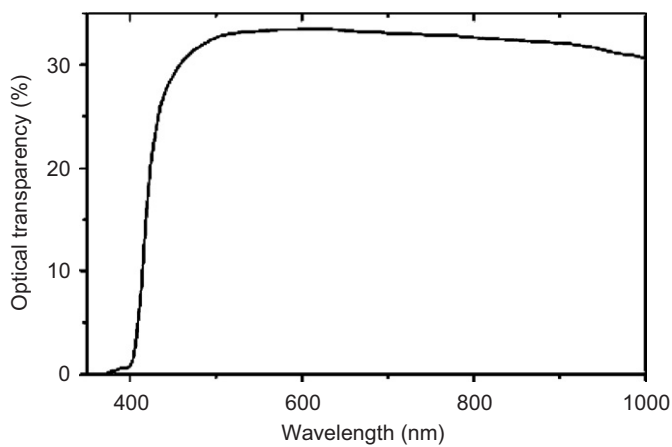

Fig. 1. Optical transparency of the ZnO:Zn ceramics of 1.6 mm thickness. Inset: a photo of a fragment of the ZnO:Zn.

The specimens of a 20 mm diameter and 1.6 mm thickness were used in the measurements. The concentration of excess Zn in ZnO was within $10^{18}$–$10^{19}$ cm$^{-3}$. All specimens were checked by X-ray diffraction. The density of the obtained ZnO:Zn ceramics was more than 99% of that for single crystal. The microstructure of the ceramic specimens was studied using an optical microscope.

The emission spectra were measured at a steady state X-ray excitation (40 kV, 10 mA) with a photomultiplier tube FEU-106 and a MDR-2 grating monochromator with 1200 lines/mm. The same set-up was used for measurement of TSL in the range from 80 to 500 K. The kinetic curves were recorded at pulsed (<1 ns) X-ray excitation using standard START-STOP detection system described elsewhere (Rodnyi et al., 2001). The emission spectra were measured in the reflection mode: the angle between a X-ray beam and a photomulpier was 90°. Decay time curves of the ceramics were measured in the transmission mode: a photomulpier tube was placed behind the sample in the line with a X-ray beam. The scintillation characteristics of the ZnO:Zn ceramics were compared with those of $Bi_4Ge_3O_{12}$ (BGO) scintillator of the same thickness using it as a standard. The studied sample and BGO were placed as far as possible in similar conditions.

## 3. Results

The emission spectrum of the ZnO:Zn ceramics is shown in Fig. 2 (curve 1). The spectrum displays an intense broad band peaking near 500 nm (2.5 eV) and a weak narrow band peaking at 380 nm (3.2 eV). In pure polycrystalline ZnO the similar emission bands have almost equal intensities (curve 2, Fig. 2). Emission spectrum of the BGO scintillator is presented in Fig. 2, curve 3, for comparison.

The light output, which was determined as area under emission curve, of the ZnO:Zn ceramics was 80% of that of BGO. The light output of the ZnO:Zn ceramics increases by factor 2 at cooling of the specimen to 80 K. This increase occurs owing to the growth of the blue-green emission, the intensity of the 380 nm band does not change with temperature.

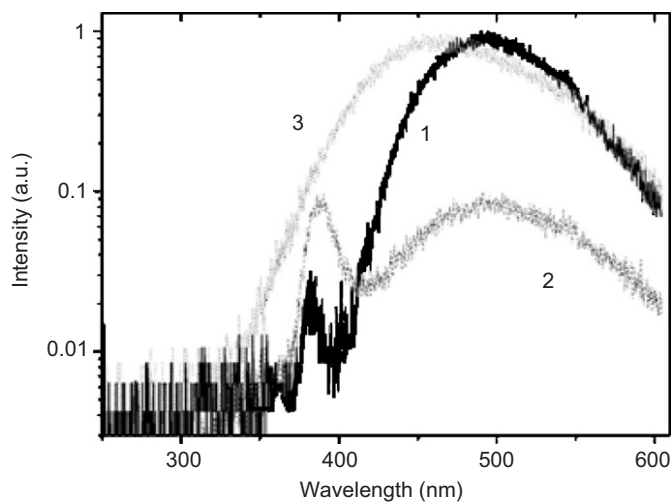

Fig. 2. X-ray induced emission spectra of the ZnO:Zn ceramics (1), the pure polycrystalline ZnO (2) and BGO (3), $T = 295$ K.

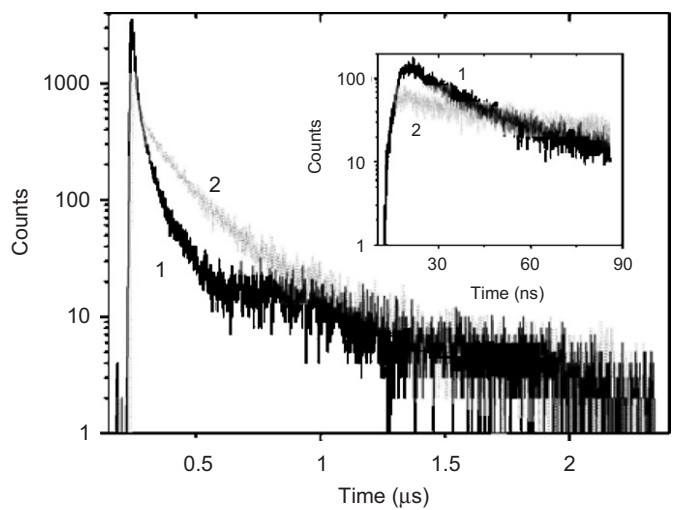

Fig. 3. Luminescence decay time curve of the ZnO:Zn ceramics (1) and BGO (2) under X-ray excitation at room temperature. Inset: the beginning parts of the curves.

The ZnO powder shows a proportional growth of both emission bands with decreasing the temperature. Besides, at 80 K the 3.2 eV band shows a structure consisting of four additional maxima with a step of about 0.1 eV between them. This value is closed to the energy of longitudinal optical phonons in ZnO, 0.073 eV (Park et al., 1966). So, we can suppose that additional maxima appear from phonon replications.

The luminescence kinetic characteristics of the ZnO:Zn in comparison with that of BGO are presented in Fig. 3. One can see that the ZnO:Zn ceramics shows better decay time characteristics than BGO. The decay curve of the ZnO:Zn can be approximated by two exponential components: fast one with constant $\tau_1 = 10.4 \pm 0.1$ ns, and slow one with constant $\tau_2 = 1.15 \pm 0.08$ μs. One can see that contribution of the fast component in the total yield is a considerable one. We anticipate that the ZnO:Zn possesses a low level of afterglow, because the scintillation intensity at 50 μs is 0.1% of the maximum intensity.

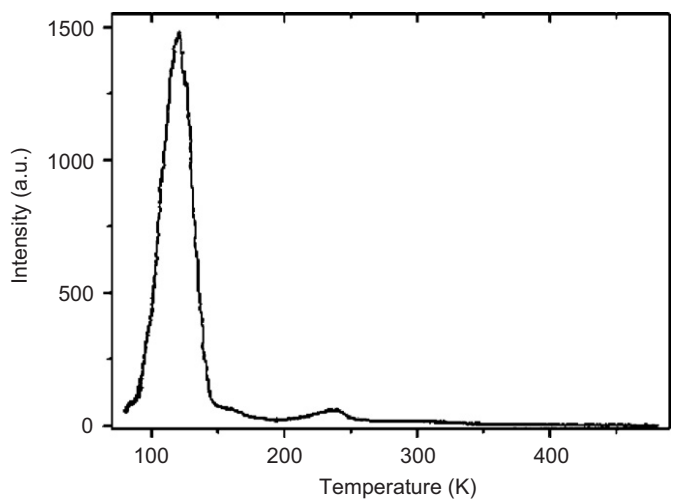

Fig. 4. TSL glow curve of the ZnO:Zn ceramics after X-ray irradiation at 80 K.

The TSL glow curve of the ZnO:Zn ceramics irradiated during 100 s by X-rays at 80 K is shown in Fig. 4.

## 4. Discussion

The observed optical transparency (33% at 1.6 mm thickness) of the ZnO:Zn ceramics fabricated by the hot pressing technique is not a bad result for a non-cubic material. The ceramics show two emission bands like ZnO:Zn and ZnO powders. A relatively week and sharp UV band (380 nm) (below absorption edge) is well studied in pure ZnO and assigned to a radiative recombination of excitons (Van Dijken et al., 2000; Wilkinson et al., 2001). The origin of a more intense and broad band (500 nm) in the visible part of the spectrum was unknown for a long time. Recently Van Dijken et al., (2000) have shown that this band arises from a recombination of electrons on oxygen vacancies. Excess interstitial Zn increases the number of oxygen vacancies in ZnO (Lott et al., 2004) and, as a result, the intensity of the visible emission grows.

The shapes of the emission spectra of the ZnO:Zn ceramics and the ZnO:Zn powder from which this ceramics has been prepared are very similar. Note, that the ZnO:Zn ceramics fabricated by spark plasma sintering (SPS) technique emits only a UV band (Kubota et al., 2004). These authors assigned the disappearance of the 500 nm emission band to a strongly reducing atmosphere of the SPS technique.

We have not carried out time-resolved spectrum measurements, but an experiment with the corresponding filters shows that the fast (∼ 10 ns) and slow (∼ 1 μs) components of the X-ray stimulated emission are connected with the UV and visible bands, respectively. The radiative lifetime of excitons in pure ZnO lies in the subnanosecond range (Wilkinson et al., 2001), while in the ZnO:Zn ceramics the observed decay of the excitonic band is about 10 ns. The increase in the decay time of the ZnO:Zn may be related to the presence of shallow traps, which are formed due to the boundaries between grains of ceramics. Thus, in the ZnO:Zn ceramics obtained from the standard ZnO:Zn phosphor the excitonic emission (380 nm band) produces fast scintillations, while the slow component (500 nm emission band) is responsible for a high light output. The concentration of excess Zn in the ZnO:Zn phosphor was chosen from the consideration of maximum light yield. So, we suppose that if scintillation response time were more important than its light output, one could increase the contribution of the fast component of the ZnO:Zn ceramics via the Zn concentration decrease.

The result of TSL measurement shows that the depth of main traps or defects created at irradiation of the ZnO:Zn ceramics at liquid nitrogen temperature is 0.21 eV. According to Lott et al. (2004) the shallow traps in ZnO are associated with the residual hydrogen impurity. It is worth noting that the irradiation of the ZnO:Zn ceramics at room temperature does not produce a noticeable TSL which is an evidence of high radiation hardness of the studied specimen.

## 5. Conclusions

The optical ceramics ZnO:Zn of 20 mm diameter and 1.6 mm thickness have been prepared by hot pressing technique. A main wide emission band peaking at 500 nm is presumably associated with the recombination of electrons on oxygen vacancies. This band provides the high light output of the ZnO:Zn ceramics. The 380 nm narrow band, which is caused by excitonic emission, is responsible for fast response of the scintillator. In general, the obtained ZnO:Zn ceramics offer rather good characteristics:

- optical transparency of up to 33% in the visible region;
- high intensity of the X-ray stimulated luminescence, which is comparable with that for standard BGO scintillator;
- main decay constant of about 10 ns;
- low level of afterglow;
- predominant blue-green emission;
- high radiation hardness.

The elaborated hot pressing method can be used for forming other ZnO-based ceramic scintillators from powder compounds. Finally, we expect that ZnO-based ceramics can become potential candidates for new fast scintillators.